\documentstyle[12pt]{article}
\catcode`\@=11
\renewcommand{\thefootnote}{\fnsymbol{footnote}}

\begin{document}

\begin{titlepage}

{\hfill DFPD/93/TH/69}

\vspace{0.2cm}

{\hfill hept-th/9403079}

\vspace{1.2cm}

{\centerline{{\bf THE HIGGS MODEL FOR ANYONS AND LIOUVILLE
ACTION:}}}

\vspace{0.5cm}

{\centerline{{\bf CHAOTIC SPECTRUM, ENERGY GAP AND EXCLUSION
PRINCIPLE}\footnote[5]{Partly Supported by the European Community Research
Programme {\it Gauge Theories, applied supersymmetry and quantum
gravity}, contract SC1-CT92-0789}}}

\vspace{1.6cm}

\centerline{\large{\sc Marco}  {\sc  Matone}\footnote{e-mail:
matone@padova.infn.it, mvxpd5::matone}}

\vspace{0.1cm}

\centerline{\it Department of Physics ``G. Galilei'' -
Istituto Nazionale di Fisica Nucleare}
\centerline{\it University of Padova}
\centerline{\it Via Marzolo, 8 - 35131 Padova, Italy}

 \vspace{2.1cm}

\centerline{\bf ABSTRACT}

\vspace{0.2cm}

Geodesic completness and self-adjointness
imply that the Hamiltonian for anyons is the Laplacian
with respect to the Weil-Petersson metric. This
metric is complete on the Deligne-Mumford compactification
of moduli (configuration) space.
The structure of this compactification fixes
the possible anyon configurations.
This allows us to identify anyons with singularities
(elliptic points with ramification
$q^{-1}$) in the Poincar\'e  metric implying
that anyon spectrum is chaotic for $n\ge 3$.
Furthermore, the bound on the holomorphic sectional curvature of moduli
spaces implies a gap in the energy spectrum.
For $q=0$ (punctures)
anyons  are infinitely separated in the Poincar\'e metric (hard-core).
This indicates that the exclusion principle has a geometrical
intepretation.
Finally we give the differential equation satisfied by the generating
function for volumes of the configuration space of anyons.

\end{titlepage}

\newpage

\setcounter{footnote}{0}

\renewcommand{\thefootnote}{\arabic{footnote}}

\noindent
{\bf 1.} Let us begin with the following remark

\begin{itemize}

\item[{\bf a}.]{{\it The configuration space of $n$ anyons is the
 space of $n$ unordered points in} $\widehat{\bf C}={\bf C}\cup
\{\infty \}$
$$
 M_n=(\widehat{\bf C}^n\backslash \Delta_n)/{Symm(n)},
$$
{\it with $\Delta_n$ the diagonal subset where two or more punctures
coincide.}}
\item[{\bf b}.]{{\it The
Liouville action on the Riemann sphere with $n$-punctures evaluated on the
classical solution is the K\"ahler potential for the natural metric
(the Weil-Petersson metric) on\footnote{The $PSL(2,{\bf C})$ group
reflects the M\"obius symmetry of the Riemann sphere. Thus one can
consider ${\cal M}_n$ as the configuration space for $n-3$ anyons
on the Riemann sphere with 3-punctures.}}
$$
{\cal M}_n=
(\widehat{\bf C}^n\backslash \Delta_n)/{Symm(n)}\times
PSL(2,{\bf C}).
$$}
\end{itemize}

This remark implies that starting from anyons on $\widehat{\bf C}$ one can
recover the two-form associated to the
natural metric on the configuration space by first computing
the Poincar\'e metric $e^{\varphi}$
on the punctured sphere and then, after evaluating
the Liouville action for $\phi=\varphi$, computing the
curvature two-form of the Hermitian line bundle on ${\cal M}_n$
defined by the classical action (see below).

To understand the physical relevance of this remark
we notice that the quantum Hamiltonian for $n$ anyons is
proportional to the covariant Laplacian on $M_n$. In the case of anyons
on the thrice punctured Riemann sphere we have
\begin{equation}
H=-{1\over 2m}\Delta_{WP},
\label{quanhamil}\end{equation}
where $\Delta_{WP}$ is the Weil-Petersson Laplacian.

A crucial point is the
physical requirement of self-adjointness of the Hamiltonian.
In our  approach this requirement
is satisfied
by considering the Hamiltonian defined on $\overline{\cal M}_n$ where
$\partial\overline{\cal M}_n=\overline{\cal M}_n\backslash {\cal M}_n$
denotes the Deligne-Mumford
compactification of ${\cal M}_n$.
In particular we recall that the Weil-Petersson metric defines
a Hermitian inner product
and has a completion to the boundary $\partial\overline{\cal M}_n$.
Thus
(\ref{quanhamil}) describes a well-defined physical problem.
In particular the structure of the Deligne-Mumford
compactification fixes possible anyon-configurations.
We notice that in the
degeneration limit punctures never collide (hard-core).
This is a consequence of the requirement
of stability.
 On the other hand this can be seen as a consequence
of the fact that the Liouville equation has
delta-singularities at the punctures (hard-core).

The connection between anyons and Liouville arises also
in considering the  critically coupled Abelian Higgs model in
(2+1)-dimensions where the space $M_n$
plays a crucial role in the analysis of the $n^{th}$ topological sector of the
theory.
Remarkably $f=2{\rm Re}\,\log \phi$, with
$\phi$ the Higgs field, satisfies the modified
(non covariant) Liouville equation
\begin{equation}
f_{z\bar z}=e^f-1,
\label{oid56}\end{equation}
which is the second Bogomol'nyi equation.
Notice that the non covariance (the $-1$
in (\ref{oid56})) is due to the Higgs mass.
In \cite{samols} Samols has studied the metric on vortex moduli space
in the Abelian Higgs model. In \cite{manton} Manton has proposed a model for
anyons which is based on the structure of such a space.
The Lagrangian density of the Abelian Higgs model is
\begin{equation}
{\cal L}={1\over 2}D_\mu \phi {\overline {D^\mu \phi}}-
{1\over 4}F_{\mu\nu}F^{\mu\nu}-{1\over
8}\left(|\phi|^2-1\right)^2,
\qquad \phi=\phi_1+i\phi_2,
\label{higgs34}\end{equation}
where $D_\mu\phi=(\partial_\mu-iA_\mu)\phi$, $F_{\mu\nu}=
\partial_\mu A_\nu-\partial_\nu A_\mu$,
and the metric has signature $(1,-1,-1)$.
In the temporal gauge
the finiteness of the energy implies that at infinity
the Higgs field is a pure phase.
The magnetic flux through ${\bf R}^2$ is
$\int F_{12}=2\pi n$
where $n$ is the winding number labelling the topological sectors
of the map
$|\phi|: S^1_\infty\longrightarrow U(1)$.
 In the static
configuration $\dot A_i=0$, $\dot \phi=0$, the energy has the
lower bound $E\ge \pi |n|$. The critical case $E=\pi |n|$ arises
when the Bogomol'nyi equations
\begin{equation}
\left(D_1+{\rm sgn}(n) i D_2\right)\phi=0,
\qquad F_{12}+{\rm sgn}(n){1\over 2}\left(|\phi|^2-1\right)=0,
\label{bogomolnyi}\end{equation}
are satisfied.
It is crucial that in the $n^{th}$ topological sector the space of smooth
solutions is a
manifold $\widetilde M_n$ of complex dimension $n$.
In particular, each solution is uniquely specified by the $n$
unordered points $\{z_k\}$ where the Higgs field is zero
\cite{jt}. The same happens in considering the Liouville equation
for the Poincar\'e metric on Riemann surfaces. In particular,
due to the uniqueness of the solution of the Poincar\'e metric,
to each complex structure of the $n$-punctured Riemann sphere
(the unordered set of $n$ points) corresponds a solution of the
Liouville equation (see below for details).

Topologically
$\widetilde M_n$ coincides with $\widehat{\bf C}^n$.
It is a remarkable fact that the Abelian Higgs model (almost) provides
a smooth metric and $U(1)$ gauge field on $\widetilde M_n$.
In particular the kinetic energy induces the metric
$$
ds^2={1\over 2}\int_{\bf C}(d\phi_a d\phi_a +dA_idA_i).
$$
The field evolution is described by geodesic motion on $\widetilde M_n$.
Notice that this corresponds to deformation of the complex structure
of punctured spheres.
The $z_k$'s are
good coordinates only on the subspace $M_n$.
Good global coordinates on $M_n$ are provided by
the coefficients of the polynomial \cite{h}
$P_n(z)=\sum_{k=0}^nw_kz^k\equiv \prod_{k=1}^n (z-z_k)$.
We stress that the choice of good coordinates is a subtle point.
Actually, as we will see, instead of working with this kind of structure
it is much more convenient to work with the Weil-Petersson metric.
This is directly related to Liouville theory and Fuchsian groups. This
structure is at the origin of chaotic spectra for anyons.

In \cite{samols} Samols introduced the metric
$ds^2={1\over 2}\sum_{r,s=1}^n
\left(\delta_{rs}+2{\partial \bar b_r\over \partial z_s}\right)dz_rd
\bar z_s$ on $\widetilde M_n$
where the $b_r$'s satisfy the equations
\begin{equation}
{\partial b_r\over \partial \bar z_s}={\partial \bar b_s\over
\partial z_r}.\label{accsrprlike}\end{equation}

\vspace{1cm}

\noindent
{\bf 2.} Let us denote by $H$ the upper half-plane and with $\Gamma$ a finitely
generated Fuchsian group. A Riemann surface isomorphic to the quotient
$H/\Gamma$ has the Poincar\'e metric $\hat g$ as the unique metric with scalar
curvature $R_{\hat g}=-1$ compatible with its complex structure. This implies
the uniqueness  of the solution of the Liouville equation on $\Sigma$
\begin{equation}
\partial_z\partial_{\bar {z}}\varphi(z,\bar{z})=
{1\over 2}e^{ \varphi(z,\bar{z})}.\label{1}\end{equation}
Let us consider the punctured Riemann sphere $\Sigma=\widehat
{\bf C}\backslash\{z_1,\ldots,z_n\}$,
$\widehat {\bf C}\equiv {\bf C}\cup\{\infty\}$.
The moduli space of $n$-punctured spheres is the
space of classes of isomorphic $\Sigma$'s, that is
\begin{equation}
{\cal M}_{n}=
\{(z_1,\ldots,z_{n})\in
\widehat{\bf C}^{n}|z_j\ne z_k\; {\rm for}\; j\ne k\}/Symm(n)\times
PSL(2,{\bf C}), \label{modulisp}\end{equation}
where ${Symm}(n)$ acts by permuting
$\{z_1,\ldots,z_n\}$ whereas $PSL(2,{\bf C})$ acts by linear fractional
transformations. By $PSL(2,\bf C)$ we can recover the `standard
normalization':  $z_{n-2}=0$, $z_{n-1}=1$ and $z_{n}=\infty $.
  For the classical Liouville tensor $T^F=\varphi_{zz}-\varphi_z^2/2$ we have
$$
T^F(z)
=\sum_{k=1}^{n-1}\left({1\over 2(z-z_k)^2}+
{c_k\over z-z_k}\right),\qquad
\lim_{z\to \infty}T^F(z)={1\over 2z^2}+{c_n\over z^3}+
{\cal O}\left({1\over |z|^4}\right).
$$
The $c_k$'s, called accessory parameters,
are functions on the space
\begin{equation} V^{(n)}=\{(z_1,\ldots,z_{n-3})\in
{\bf C}^{n-3}|z_j\ne 0,1; z_j\ne z_k,\; {\rm for}\; j\ne k\},
\label{8}\end{equation}
which is a covering of ${\cal M}_{n}$
\begin{equation}
{\cal M}_{n}\cong V^{(n)}/{Symm}(n),
\label{mdls}\end{equation}
where the action of $Symm(n)$ on $V^{(n)}$ is defined by comparing
(\ref{modulisp}) with (\ref{mdls}).

The Liouville
action on the Riemann spheres with $n$-punctures
$$
S^{(n)}=
\lim_{r\to 0}\left[\int_{\Sigma_r}
\left(\partial_z\phi\partial_{\bar z}{\phi}+e^{\phi}\right)+
2\pi (n {\log} r+2(n-2){\log}|{\log}r|)\right],
$$
$\Sigma_r=\Sigma\backslash\left(\bigcup_{i=1}^{n-1}
\{z||z-z_i|<r\}\cup\{z||z|>r^{-1}\}\right)$
has the following property \cite{0}
 \begin{equation}
-{1\over 2\pi}{\partial S^{(n)}_{cl}\over \partial z_k}=c_k,
\qquad k=1,\ldots,n-3,
\label{33}\end{equation}
where the $c_k$'s are the accessory parameters.
Since $S^{(n)}_{cl}$ is real, eq.(\ref{33}) yields
${\partial c_j\over \partial {\bar z_k}}=
{{\partial \bar c_k\over \partial {z_j}}}$ corresponding to the
Samols equations
(\ref{accsrprlike}).
It turns out that \cite{0}
 \begin{equation}
{\partial c_j\over\partial\bar z_k}={1\over 2\pi}\left \langle {\partial\over
\partial z_j}\,,{\partial\over \partial z_k} \right \rangle^{(n)}, \qquad
j,k=1,\ldots,n-3,\label{34}\end{equation}
where the brackets denote
the Weil-Petersson metric on the Teichm$\ddot{\rm u}$ller space
$T_{n}$ projected onto $V^{(n)}$.
Therefore by (\ref{33})
\begin{equation}\omega_{WP}^{(n)}= {i\over
2}{\overline\partial}{\partial}S^{(n)}_{cl}=-i\pi\sum_{j,k=1}^{n-3}
 {\partial c_k\over \partial {\bar z_j}}d\bar z_j\wedge d z_k,
\label{36}\end{equation} where $\omega_{WP}$ is the Weil-Petersson
two-form  on $V^{(n)}$.

\vspace{1cm}

\noindent
{\bf 3.} The symplectic structure above allows us to consider
the Poisson bracket relations \cite{t2}
\begin{equation}
\{c_j,c_k\}_{WP}=0,\qquad \{c_j,z_k\}_{WP}={i\over \pi}\delta_{jk},
\label{bb1aaa}\end{equation} where the brackets are defined with
respect to $\omega_{WP}$.
These relations suggest performing
 the geometric quantization of space ${\cal M}_n$.
To do this we must define a suitable
line bundle. Let $\sigma_{k,n}$ be the element of $Symm(n)$ interchanging
$k$ and $n$. Let us consider the function \cite{asym}
\begin{equation}
f_{\sigma_{k,n}}={
\prod_{j\ne k,\,j=1}^{n-3}(z_j-z_k)^2\over (z_k(z_k-1))^{n-4}}, \;
k=1,\ldots, n-3,\quad f_{\sigma_{n-2,n}}=\prod_{j=1}^{n-3}z_j^2,\quad
f_{\sigma_{n-1,n}}=\prod_{j=1}^{n-3}(z_j-1)^2.
\label{ccycl1}\end{equation}
Extension by the composition
$f_{\sigma_1\sigma_2}=(f_{\sigma_1}\circ \sigma_2)f_{\sigma_2}$
defines a 1-cocycle $\{f_\sigma\}_{\sigma\in{Symm}(n)}$ of
${Symm}(n)$ \cite{asym}.
Let us consider the holomorphic line bundle
${\cal L}_{n} =V^{(n)}\times {\bf C}/{Symm}(n)$,
 on ${\cal M}_{n}$  where the action of $\sigma\in
{Symm}(n)$ is defined by
$(x,z)\to (\sigma x,f_\sigma (x)z)$, $x\in V^{(n)}$, $z\in {\bf C}$.
Since
$\exp \left({ S^{(n)}_{cl}\circ \sigma
 / \pi}\right)|f_\sigma|^2=\exp \left({ S^{(n)}_{cl}
/\pi}\right)$
 it follows that
$\exp (S^{(n)}_{cl}/\pi)$ is a
  Hermitian metric in the line bundle ${\cal L}_{n}\to
{\cal M}_{n}$.
By (\ref{33})
 $\exp (S^{(n)}_{cl}/\pi)$ has connection form $-2\sum_i c_i d z_i$ and,
by (\ref{36}), curvature two-form
$-{ 2i\over  \pi}\omega_{WP}$ \cite{asym}.
The covariant derivatives are
\begin{equation}
{\cal D}_k =\partial_{z_k}-\partial_{z_k}{S^{(n)}_{cl}\over\pi}=
\partial_{z_k}+2c_k,\qquad
 \qquad {\overline{\cal D}}_k ={\partial}_{\bar z_k}.
\label{covq}\end{equation}
In the geometric quantization the Hilbert states
are sections of ${\cal L}_{n}$ annihilated by `half' of
the derivatives (polarization). The natural choice is
\begin{equation}
{\cal H}=\left\{ \psi\in {\cal L}_{n}| {\overline{\cal D}}_k
\psi=0\right\},
\label{hilbert1}\end{equation}
with inner product
\begin{equation}
\big<\psi_1|\psi_2\big>=
 {1\over n!(n-3)!}\int_{{\cal M}_n}
d(WP)
e^{-{S^{(n)}_{cl}\over \pi}}\overline \psi_1\psi_2,
\label{inner1}\end{equation}
where
\begin{equation}
d(WP)\equiv\left(\bigwedge^{n-3}{i\over 2}
\overline\partial\partial
S^{(n)}_{cl}\right),\label{wpv}\end{equation}
that by (\ref{36}) is the Weil-Petersson volume form.
An alternative to (\ref{wpv})
is to use the volume form $\bigwedge^{n-3} \omega$ where
\begin{equation}
\omega={i\over 2}\overline\partial\partial \log \det\,\left\|{\partial^2
S^{(n)}_{cl}\over \partial \bar z_j \partial z_k}\right\|.
\label{altern99}\end{equation}
However, in general, the correspondence principle can be proved only
if the scalar product is defined with respect to the
volume form associated with the
K${\rm \ddot a}$hler  potential \cite{berezin}. This means that in our
case we must use the Weil-Petersson volume form.

We conclude the discussion on the geometric quantization of ${\cal M}_n$
by noticing that an interesting alternative for the polarization choice in
the geometric quantization above is
\begin{equation}
{\cal H}=\left\{\widetilde \psi\in {\cal L}_{n}|
{\cal D}_k\widetilde \psi=0\right\}.\label{polarization2}\end{equation}
In this case the  states and the classical Liouville action are related by
\begin{equation}
\partial_{z_k}{S^{(n)}_{cl}}=\pi\partial_{z_k}\log
\widetilde\psi.\label{00}\end{equation}

\vspace{1cm}

\noindent
{\bf 4.}
A crucial aspect that should be further investigated concerns the
classical-quantum interplay arising both in Liouville and
anyon theories\footnote{Vortices are
unlabelled particles not
only quantum mechanically but also classically \cite{manton}.}.
In \cite{mma,mmb} it has been emphasized
that the regularization arising at the classical level
for the Liouville action is strictly related to the conformal properties
both of quantum and classical (Poincar\'e metric) Liouville operators.

The approach considered here
 makes it possible to define
 the conformal weight associated to anyons \cite{mmb}
\begin{equation}
\Delta(q)={1-q^2\over 2h},
\label{iocd}\end{equation}
where $q^{-1}$ is the ramification index of the anyon (elliptic point).
For $q=0$ the point is parabolic (puncture). In this case the geodesic
distance between an arbitrary point and an anyon is infinity in the
natural (Poincar\'e) metric. In other words, if $z_k$ is the coordinate
of the anyon (with index $q=0$) then
\begin{equation}
\lim_{Q\to z_k} d(P,Q)=\infty.
\label{gdye}\end{equation}
In particular two anyons with $q=0$ are infinitely separated!
This means that for spin $1/2h$, particles cannot coalesce in the
Poincar\'e metric (hard-core). This connection between spin and
hard-core resembles the Pauli exclusion principle. Actually this
hard-core is encoded in our approach. The reason is that the
Liouville equation on the sphere with the
punctures filled-in has additional terms
given by $\delta$-functions at the punctures.
These aspects should be useful in investigating the standard
approach to anyon \cite{fm}.

\vspace{1cm}

\noindent
{\bf 5.} We now consider  the energy spectrum of anyons.
As we have seen Liouville theory arises in the framework of negatively
curved surfaces. This suggests to consider the Gutzwiller model
\cite{GBV} for chaos.
To see this notice that in the case
of one anyon moving on the thrice-punctured sphere
the Hamiltonian (\ref{quanhamil}) reduces
to the Laplacian on the Riemann sphere with punctures fixed at
$0,1$ and $\infty$.  Note that
 fixed puntures may be considered as infinitely massive anyons.
The crucial point is that in this case the Hamiltonian corresponds to
the Laplacian on a Riemann surface of negative curvature.
Therefore
we  can apply the results in
\cite{GBV} concerning quantum chaos on Riemann surfaces.

This result can be extended to the general situation of multi-anyon
spectra. The point is that moduli space of higher genus (i.e. $h\ge 2$)
is negatively curved. In particular the holomorphic sectional curvatures
are bounded away from zero by a negative constant \cite{Royden}.
Therefore it seems that the previous result about chaoticity for
anyons is a general situtation which holds for $n\ge 3$.

The previous bound on the holomorphic sectional curvature suggests that
there is a mass-gap in the anyon spectra.
This is a direct consequence of the infrared properties of negatively
curved surfaces.
Actually this is the case if one
considers  one anyon moving on the thrice punctured sphere.
In particular it is well-known that the eigenvalues of the
Laplacian have a gap in the spectrum.

Let us make further remarks
on the  anyon dynamics. The points is that it depends on the geodesic (complex)
structure. This structure is determined by the position of the anyons
themself, resulting structure is fashinating and highly nontrivial.
Actually we can consider anyons as singularities in the metric
with a self gravitational interaction.

Results from the theory
of moduli spaces allow to compute the generating function for
the volumes of configuration space of anyons (factorized by the trivial
infinity coming form the $PSL(2,{\bf C})$ symmetry).
In particular it turns out
that the generating function satisfies
the differential equation \cite{mmd}
\begin{equation}
g''={{g'}^2t-gg'+g't\over t(t-g)},
\label{51a}\end{equation}
where
\begin{equation}
g(t)=\sum_{k=3}^\infty {k(k-2)\over (k-3)!}t^{k-1}
\int_{\overline{\cal M}_{k}}
\left({i\overline\partial\partial
S_{cl}^{(k)}\over 2\pi^2}\right)^{k-3},
\label{prtbtv}\end{equation}
where ``${\int_{\overline{\cal M}_{3}}1}$''$\equiv {1\over 6}$.

Another interesting aspect is the connection between Liouville
and Higgs. This provides a way to consider the Higgs
model in a 2D gravity framework.

Finally we stress that the relation between Liouville
theory and anyons arises also in a different context (see
\cite{Jackiw} and references therein).

\vspace{1cm}

\noindent
{\bf Acknowledgements.} I would like to thank C. Grosche for remarks
on chaos, R. Jackiw
for bringing ref.\cite{Jackiw} to my attention and P.A. Marchetti for
stimulating discussions.

\end{document}